\documentclass{IOS-Book-Article}

\usepackage{mathptmx}
\usepackage{graphicx}

\def\hb{\hbox to 10.7 cm{}}

\begin{document}

\pagestyle{headings}
\def\thepage{}

\begin{frontmatter}              % The preamble begins here.

%\pretitle{Pretitle}
\title{Extreme Scale-out SuperMUC Phase 2 - lessons learned}

%\markboth{}{September 2015\hb}
%\subtitle{Subtitle}

\author[A]{\fnms{Nicolay} \snm{Hammer}%
\thanks{Corresponding Author. E-mail: nicolay.hammer@lrz.de.}},
\author[A]{\fnms{Ferdinand} \snm{Jamitzky}},
\author[A]{\fnms{Helmut} \snm{Satzger}},
\author[A]{\fnms{Momme} \snm{Allalen}},
\author[A]{\fnms{Alexander} \snm{Block}},
\author[A]{\fnms{Anupam} \snm{Karmakar}},
\author[A]{\fnms{Matthias} \snm{Brehm}},
\author[A]{\fnms{Reinhold} \snm{Bader}},
\author[A]{\fnms{Luigi} \snm{Iapichino}},
\author[A]{\fnms{Antonio} \snm{Ragagnin}},
\author[A]{\fnms{Vasilios} \snm{Karakasis}},
\author[A]{\fnms{Dieter} \snm{Kranzlm\"uller}},
\author[A]{\fnms{Arndt} \snm{Bode}},
\author[A]{\fnms{Herbert} \snm{Huber}},
\author[B]{\fnms{Martin} \snm{K\"uhn}},
\author[B]{\fnms{Rui} \snm{Machado}},
\author[B]{\fnms{Daniel} \snm{Gr\"unewald}},
\author[C]{\fnms{Philipp V.~F.} \snm{Edelmann}},
\author[C]{\fnms{Friedrich K.} \snm{R\"opke}},
\author[D]{\fnms{Markus} \snm{Wittmann}},
\author[D]{\fnms{Thomas} \snm{Zeiser}},
\author[E]{\fnms{Gerhard} \snm{Wellein}},
\author[F]{\fnms{Gerald} \snm{Mathias}},
\author[F]{\fnms{Magnus} \snm{Schw\"orer}},
\author[F]{\fnms{Konstantin} \snm{Lorenzen}},
\author[G]{\fnms{Christoph} \snm{Federrath}},
\author[H]{\fnms{Ralf} \snm{Klessen}},
\author[I]{\fnms{Karl-Ulrich} \snm{Bamberg}},
\author[I]{\fnms{Hartmut} \snm{Ruhl}},
\author[J]{\fnms{Florian} \snm{Schornbaum}},
\author[K]{\fnms{Martin} \snm{Bauer}},
\author[K]{\fnms{Anand} \snm{Nikhil}},
\author[K]{\fnms{Jiaxing} \snm{Qi}},
\author[K]{\fnms{Harald} \snm{Klimach}},
\author[L]{\fnms{Hinnerk} \snm{St\"uben}},
\author[M]{\fnms{Abhishek} \snm{Deshmukh}},
\author[M]{\fnms{Tobias} \snm{Falkenstein}},
\author[N]{\fnms{Klaus} \snm{Dolag}},
and
\author[N]{\fnms{Margarita} \snm{Petkova}}

%\runningauthor{B.P. Manager et al.}
\address[A]{Leibniz-Rechenzentrum, Garching}
\address[B]{CCHPC - Fraunhofer ITWM, Kaiserslautern}
\address[C]{Heidelberger Institut f\"ur Theoretische Studien, Heidelberg}
\address[D]{Erlangen Regional Computer Center (RRZE), University of Erlangen-N\"urnberg, Erlangen}
\address[E]{Department of Computer Science, University of Erlangen-N\"urnberg}
\address[F]{Lehrstuhl f\"ur Biomolekulare Optik, Ludwig-Maximilians-Universit\"at M\"unchen, M\"unchen}
\address[G]{Research School of Astronomy and Astrophysics, The Australian National University, Canberra}
\address[H]{Universit\"at Heidelberg, Zentrum f\"ur Astronomie, Institut f\"ur Theoretische Astrophysik, Heidelberg}
\address[I]{Chair for Computational and Plasma Physics at the LMU, Munich}
\address[J]{Chair for System Simulation, University of Erlangen-N\"urnberg, Erlangen}
\address[K]{Chair of Simulation Techniques \& Scientific Computing, University Siegen}
\address[L]{Universit\"at Hamburg, Zentrale Dienste, Hamburg}
\address[M]{Fakult\"at f\"ur Maschinenwesen, Institut f\"ur Technische Verbrennung, RWTH Aachen University, Templergraben 64, Aachen}
\address[N]{Universit\"ats-Sternwarte M\"unchen, M\"unchen}

\begin{abstract}
We report lessons learned during the friendly user block operation 
period of the new system at the Leibniz Supercomputing Centre (SuperMUC Phase 2).
\end{abstract}

\begin{keyword}
Supercomputing\sep HPC\sep Extreme Scaling\sep MPI\sep Application Scaling
\end{keyword}
\end{frontmatter}
\markboth{Extreme Scale-out SuperMUC Phase 2 - lessons learned\hb}{Extreme Scale-out SuperMUC Phase 2 - lessons learned\hb}

\section*{Introduction}
In spring 2015, the Leibniz Supercomputing Centre (Leibniz-Rechenzentrum, LRZ), installed their new Peta-Scale System SuperMUC Phase2. Selected users were invited for a 28 day extreme scale-out block operation during which they were allowed to use the full system for their applications. 

The following projects participated in the extreme scale-out workshop:
BQCD (Quantum Physics; M. Allalen), SeisSol (Geophysics, Seismics; S. Rettenberger, A. Breuer),
GPI-2/GASPI (Toolkit for HPC; M. K\"uhn, R. Machado, D. Gr\"unewald),
Seven-League Hydro (Computational Fluid Dynamics; P. Edelmann),
ILBDC (Lattice Boltzmann CFD; M. Wittmann, T. Zeiser),
Iphigenie (Molecular Dynamics; G. Mathias, M. Schw\"orer, K. Lorenzen),
FLASH (CFD; C. Federrath, L. Iapichino),
GADGET (Cosmological Dynamics; K. Dolag, M. Petkova),
PSC (Plasma Physics; K. Bamberg), 
waLBerla (Lattice Boltzmann CFD; F. Schornbaum, M. Bauer),
Musubi (Lattice Boltzmann CFD; A. Nikhil, J. Qi, H. Klimach),
Vertex3D (Stellar Astrophysik; T. Melson, A. Marek),
CIAO (CFD; A. Deshmukh, T. Falkenstein), and
LS1-Mardyn (Material Science; N. Tchipev).

The projects were allowed to use the machine exclusively during the 28 day period, which corresponds to a total of 63.4 million core-hours, of which 43.8 million core-hours were used by the applications, resulting in a utilization of 69\%. The top 3 users were using 15.2, 6.4, and 4.7 million core-hours, respectively.

\section{System Description}

The new PetaScale System SuperMUC Phase2 consists of 6 Islands of Lenovo’s NeXtScale nx360M5 WCT system each with 512 nodes. Each node contains 2 Intel Haswell Xeon E5-2697v3 processors with 28 cores and 64 GB RAM . The compute nodes are connected via an Infiniband FDR14 network with a non-blocking intra-island and a pruned 4:1  inter-island tree topology. The complete system contains 86,016 cores  and 194 TB RAM. The double precision LINPACK Performance was measured as 2.81 PetaFlop/s. Attached to the compute nodes is a parallel filesystem based on IBM’s GPFS with 15 PetaBytes. The system runs Novell’s SUSE Linux Enterprise Edition 11, IBM’s LoadLeveler is used as batch system, Intel's C and Fortran compiler, Intel and IBM MPI and is in operation for selected users since May 12th 2015. For an extensive system description, please see www.lrz.de/services/compute/supermuc/system\_description.

\section{Overall Observations}

During the extreme scale-out workshop the performance and system monitoring were already active and therefore some statistics can be calculated.

Each Haswell node contains 28 cores which can also be used with hyperthreading activated, resulting in a maximum of 56 tasks per node. 
This maximum number of 56 tasks per node was only tested by a single software package. 
Most of the projects were using 28 tasks per node. However, a significant amount of projects was using a hybrid approach with varying numbers of OpenMP threads on each node. The two main cases for mixed OpenMP/MPI usage was 1, 2, or 4 MPI tasks per node and 28, 14, or 7 OpenMP threads, respectively. For some codes the factors of 7 instead of powers of 2 were a challenge and required some re-adjustment.

\begin{figure}%
\includegraphics[width=0.5\textwidth]{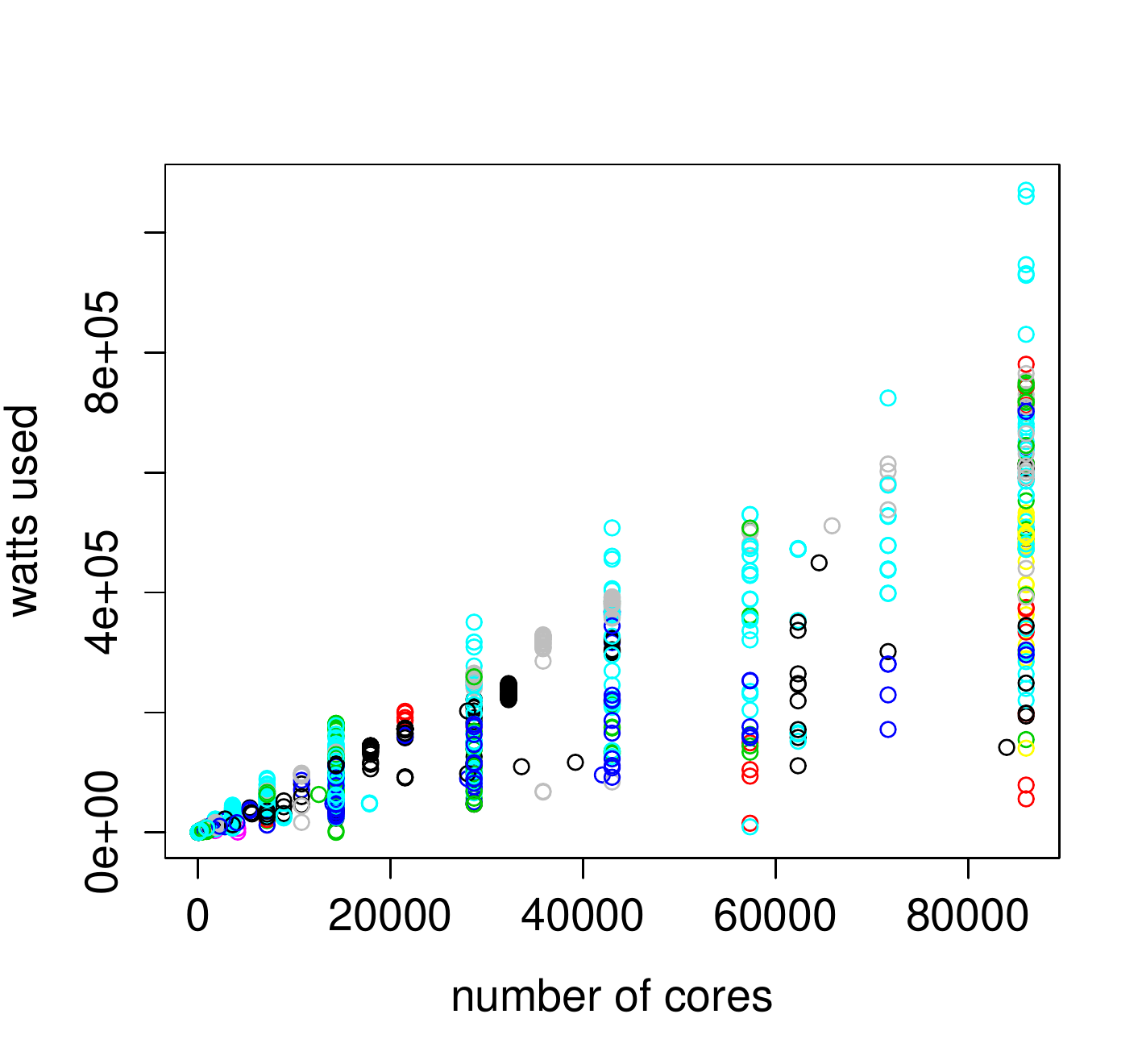}
\caption{Scaling of the total power consumption of the system versus job size,
measured by the number of cores used for
each job. Different colors decode the different software packages used.
}%
\label{fig:Rplot_taskspernode}%
\end{figure}

The schedule of the workshop was dominated by two modes of operation: special-time and general-time operation. During special-time operation mode the system was exclusivly available for a single user who could then run large jobs filling the whole system. During general-time operation mode the general queue was available where all participants of the workshop had the opportunity to submit jobs with a maximum size of half of the system. Also this operating time could be used for testing and debugging the code.

Figure \ref{fig:Rplot_taskspernode} shows the watts used for each run versus the number of cores used by this run for all jobs during the workshop. One can see that the power consumption scales nearly linear with the job size. For the largest jobs a maximum watts power of more than 1.2 MW was measured. It is crucial for the power consumption of large jobs that the node level performance is optimized. More than a relative factor of 10 in the variation of  power consumption was observed for the largest jobs with 86,016 cores which is not only important for the layout of the electrical infrastructure but also determines the power bill.

The most power hungry compute intensive simulation code SeisSol (more than 1 PFlops sustained, cyan circles) used several times more energy in comparison to the memory bound Lattice Boltzmann Code ILBDC (red circles).

\section{FLASH}
\begin{figure}
\centering\includegraphics[width=0.7\textwidth]{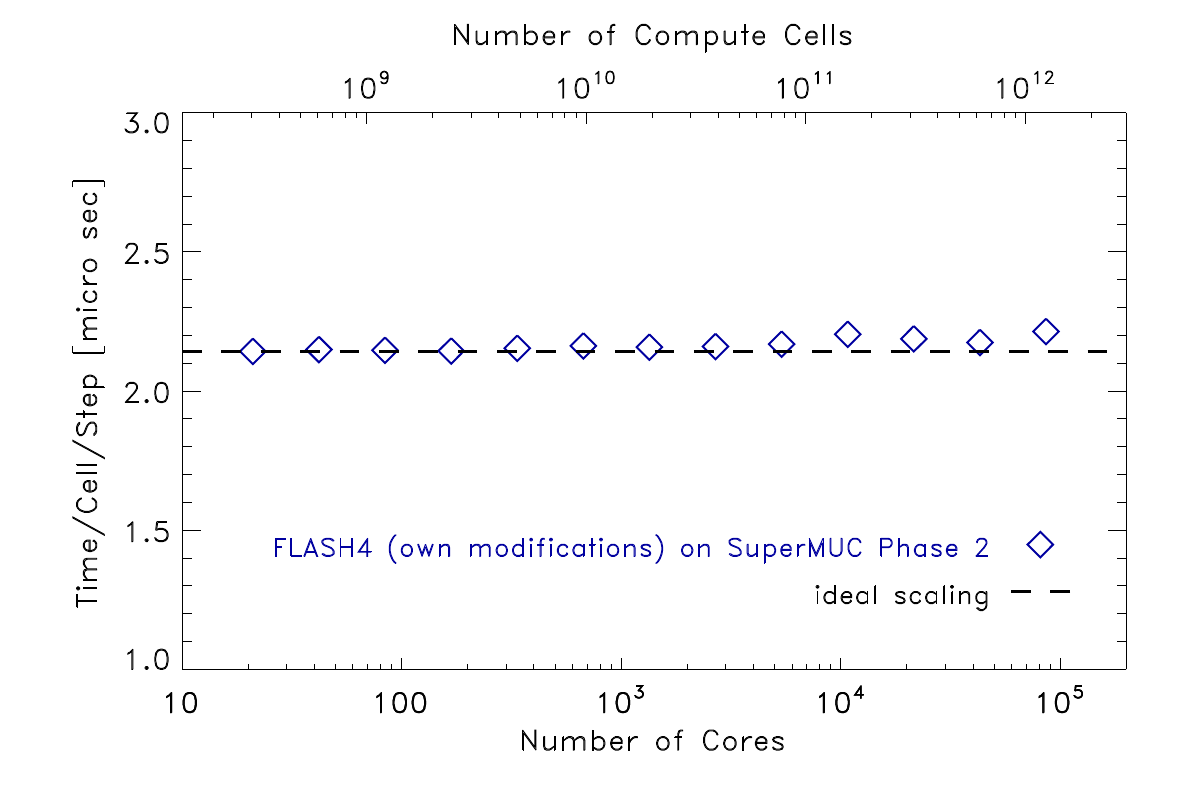}
 \caption{Weak scaling of the FLASH code.
 \label{fig:scalingtest_flash4_phase2}
 }
\end{figure}

{\sc FLASH} is a public, modular grid-based hydrodynamical code for the simulation of astrophysical flows. In the framework of the SuperMUC Phase 2 scale-out workshop, the current version (Flash4) has been optimised  to reduce the memory and MPI communication requirements. In particular, non-critical operations are now performed in single precision, without causing any significant impact on the accuracy of the results. In this way, the code runs with a factor of 4.1 less memory and 3.6 times faster than the version used for the previous large-scale project at LRZ \cite{f13}, and scales remarkably well up to the whole Phase 2 system (Figure \ref{fig:scalingtest_flash4_phase2}).

This scale-out workshop has been used for the preparation of a simulation of supersonic, isothermal turbulence with an unprecedented resolution exceeding $10,000^3$ grid elements. The simulation is expected to require several million core-hours; it will use about $155\ {\rm TB}$ of memory, 
and every data dump will be $19\ {\rm TB}$ in size.
This simulation is devoted to studying the complex flow pattern in the interstellar medium, with the aim of fully resolving the sonic length. It marks the transition scale between compressible and incompressible turbulence where the turbulent cascade crosses from the supersonic into the subsonic regime.

\section{Seven-League Hydro}

The SLH (Seven-League Hydro) code is an astrophysical hydrodynamics code that
solves the compressible Euler equations using a finite-volume method including
source terms for gravity, nuclear reactions and thermal conduction, focusing
on the simulation of the interior of stars. As these
generally involve flows at very low Mach numbers, SLH includes special
discretizations that reduce the excessive artificial dissipation which
usually occurs in this regime (see \cite{miczek2015a}).
It also uses implicit time-stepping to overcome the CFL criterion, which
is exceedingly restrictive at low Mach numbers, leading to an overall
more efficient scheme.  The most time-consuming part of the computation
is solving very large sparse Jacobian matrices using iterative linear
solvers (e.g.\ BiCGstab or multigrid).  This is mostly memory bandwidth
limited.  The code is parallelized with MPI and OpenMP, allowing pure
MPI and hybrid MPI+OpenMP configurations.

For the scaling tests on SuperMUC Phase 2 we chose the Taylor--Green vortex, which 
is commonly used as a benchmark for the
capability of codes to simulate turbulence. It can easily be scaled to an
arbitrary number of grid cells, has a very homogeneous flow pattern and does not
involve any source terms. We tested strong scaling on a $2,016^3$ grid up to the full machine.
The minimum number of cores needed to store the sparse Jacobian matrix
in memory is 21,504, using $\sim 2.3$~GiB/core. The code shows more
than ideal scaling behavior in the step from 21,504 to 43,008 cores. This
is very likely due to cache effects.
Similar behavior was observed on SuperMUC Phase 1. It does not show with
smaller grid sizes that use less memory per core. Hybrid and pure MPI
parallelization behave very similarly using the IBM MPI library.
Acceptable scaling is possible even with 86,016
processes using IBM MPI. Intel MPI could not be tested with a high number of MPI tasks
due to limitations in the library at the time.

%\begin{figure}
%\centering\includegraphics[width=0.5\textwidth]{slh-plot.pdf}
% \caption{Strong scaling of the Seven-League Hydro code.
% \label{fig:slh-plot.pdf}
% }
%\end{figure}

\section{ILBDC}

\begin{figure}%
\centering
%\parbox{5.1cm}{\includegraphics[width=5cm]{slh-plot.pdf}}%
%\qquad
%\begin{minipage}{5.1cm}%
\includegraphics[width=0.9\textwidth]{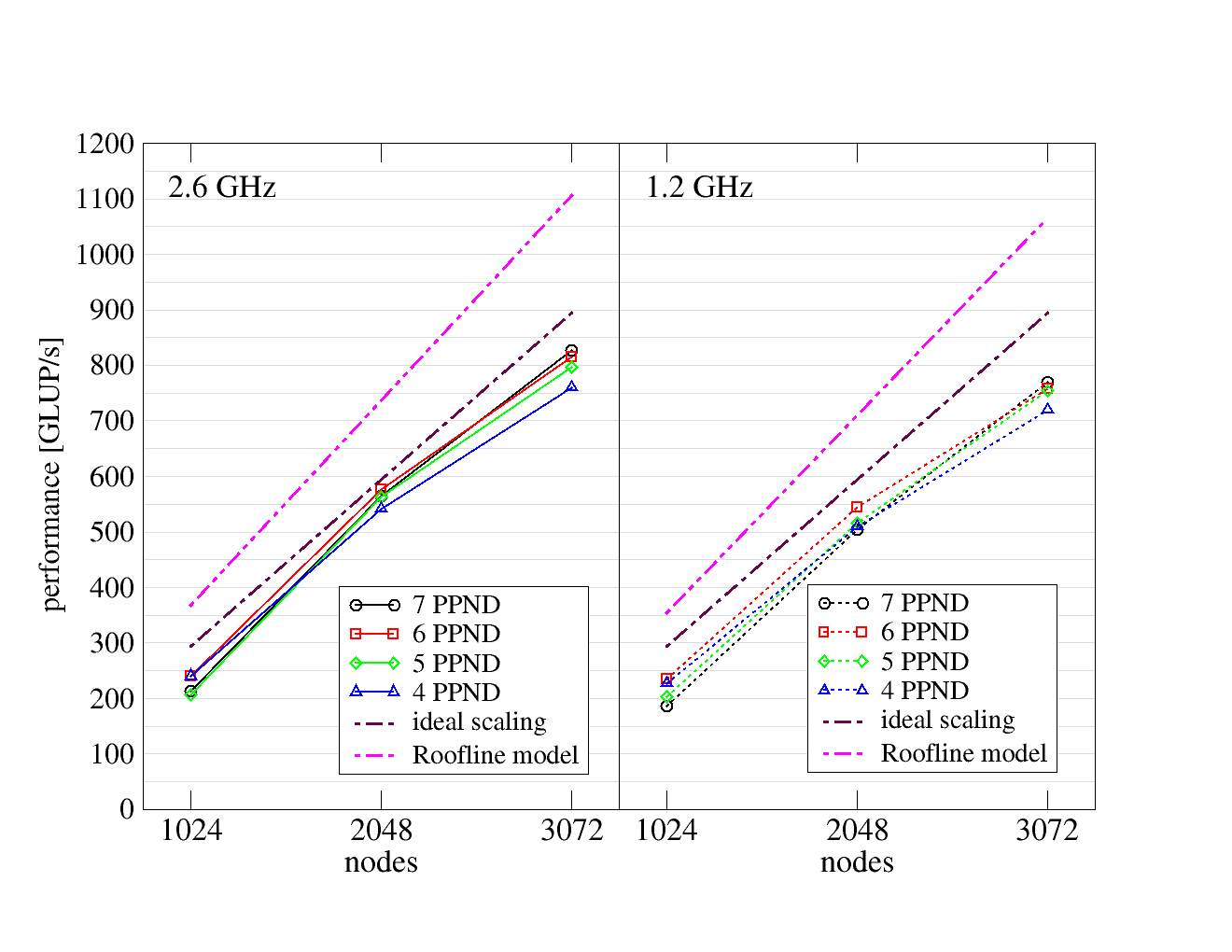}
%\end{minipage}%
\caption{Strong scaling of the ILBDC code with a fixed bed reactor geometry with $17500 \times 3500 \times 3500$ nodes at two clock frequencies
(ca.\ $10^{11}$ fluid nodes $\approx 19,6$\,TiB).
}%
\label{fig:ILBDC}%
\end{figure}

ILBDC~\cite{wittmann-2015} is a
D3Q19-TRT lattice Boltzmann flow solver.
Figure~\ref{fig:ILBDC} (right panel) shows the \textit{strong scaling} behavior
of ILBDC from two to six islands for $2.6$\,GHz (left panel).
Despite the 19 concurrent memory streams and indirect access,
ILBDC sustains $87$\,\% of the memory bandwidth achieved with the
simple STREAM copy benchmark.
Typically $5$\,PPND (processes per NUMA domain) saturate the
memory bandwidth and thus already sustained the same performance
up to $7$\,PPND.
In the large-scale case communication partially effects
this behaviour.
With the Haswell-EP architecture, the sustained memory bandwidth
of a node is almost independent of the core frequency.
Consequently ILBDC at $1.2$\,GHz still reaches $93$\% of the
performance at $2.6$\,GHz (Fig.~\ref{fig:ILBDC}, right panel).
For memory-bound codes reducing the core frequency and the number
of cores used per NUMA domain has only minimal performance impact.
This bears the potential to drastically save energy on SuperMUC
phase~2.
%\begin{figure}
%\centering\includegraphics[width=0.9\textwidth]{Smp2StrongScalingFullMachine2.png}
% \caption{
%Strong scaling of a fixed bed reactor geometry with $17500 \times 3500 \times 3500$ nodes
%(ca.\ $10^{11}$ fluid nodes $\approx 19,6$\,TiB).
% \label{fig:ilbdc}
% }
%\end{figure}

\section{Iphigenie}
The MPI/OpenMP-parallel molecular dynamics (MD)
package {\sc iphi\-genie/}{\sc cpmd}\cite{Schwoerer2015}
links the molecular mechanics (MM) C-code {\sc iphigenie} (which can be downloaded from
www.sourceforge.net/projects/iphigenie)
with the quantum--mechanical (QM) density functional theory (DFT) Fortran-implementation
{\sc cpmd} (www.cpmd.org). It is designed for highly accurate and
efficient hybrid  QM/MM--MD simulations of biomolecules in their
native solvent environment. In particular, the use of polarizable
MM force fields for the aqueous solution set it apart from
related approaches.
Efficient sampling of biomolecular conformations is
achieved by employing generalized ensemble techniques,
which jointly simulate multiple replicas of a system
and require only sparse and infrequent communication
between the replicas.

The setup for the scaling test comprises the
Alanine dipeptide (Ac-ALA-NHMe) molecule as the DFT fragment at a plane wave cutoff
of 100 Ry. It is solvated in 4500 water molecules described by
a complex polarizable water model. A total of 128 replicas of the
system span the generalized ensemble. Every node hosted 4 MPI processes
each using 7 OpenMP threads.

\begin{figure}%
%\centering
\includegraphics{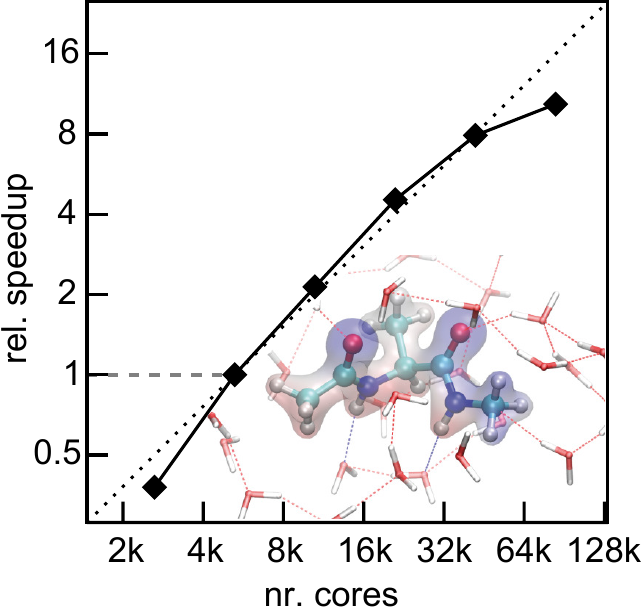}%
%\qquad
%\begin{minipage}{6.1cm}%
%\includegraphics[width=6cm]{gadget_fig1.png}
%\end{minipage}%
\caption{Strong scaling of a {\sc iphi\-genie/cpmd} generalized ensemble MD simulation
on SuperMUC Phase II for an Alanine dipeptide  (DFT) in
polarizable water (MM) as sketched in the inset.
}%
\label{fig:iffiscale}%
\end{figure}

Figure \ref{fig:iffiscale} (left panel) shows the speedup of the calculation
achieved on Phase 2 of the SuperMUC system vs.\ the number of cores employed.
Choosing the reference speed at 5376 cores,
we find perfect scaling up to half the machine and good scaling
on the full machine (86016 cores).
The graph even hints at some super-scaling
effects. Based on this excellent scaling
fundamental questions related to protein folding and vibrational
spectroscopy are now targeted on SuperMUC Phase 2
with unprecedented accuracy.

\section{The Plasma Simulation Code (PSC)}

The Plasma Simulation Code (PSC) is a general purpose framework to solve the extended Maxwell-Vlasov and extended Maxwell-Vlasov-Boltzmann equations via the Particle-In-Cell (PIC) approach. Recent extensions comprise the self-field effects of radiation and electron-positron pair production in strong fields.
The PSC is a well-tested, widely-recognized and reliable Particle-in-Cell Code that has also been fundamental to other codes, e.g. the EPOCH code.
The code was ported to a modern modularized C simulation framework supporting bindings to
FORTRAN and C/CUDA that features selectable field and particle pushers.
The framework allows the simulation of a large set of problems ranging from ultra-thin
foils as light sources and proton-driven wake fields to QED effects in extreme laser fields.
The PSC enables the simulation of large-scale, three-dimensional problems, e.g.\ ELI and AWAKE (CERN).

%\begin{figure}
%\centering\includegraphics[width=0.9\textwidth]{psc_fig_1.pdf}
% \caption{
%Strong scaling of the PSC code.
% \label{fig:psc1}
% }
%\end{figure}

%\begin{figure}%
%\centering
%\parbox{5.1cm}{\includegraphics[width=5cm]{psc_fig_1.pdf}}%
%\qquad
%\begin{minipage}{5.1cm}%
%\includegraphics[width=5cm]{tti_speedup.pdf}
%\end{minipage}%
%\caption{Left panel: Strong scaling of the PSC code.
%Right panel: Scaling of FRTM using GASPI on SuperMUC Phase 2.
%}%
%\label{fig:PSC}%
%\end{figure}

\section{GASPI}

The framework for reverse time migration (FRTM) was chosen as showcase for GPI-2. This code
is used in exploration for oil and gas (seismic imaging). It fully respects the two-way wave equation and provides
high quality imaging also in regions with complex geological structures, e.g. regions composed of steep
flanks and/or high velocity contrasts. RTM has high demands on
the underlying compute resources. Complex physical modeling, large target output domains, large
migration aperture and/or high frequency content require efficient parallelization on the algorithmic
side.

In FRTM, the parallelization is based on a fine granular domain decomposition on a regular grid. It is
supplemented by a fully asynchronous data dependency driven task based execution. Remote
completion is used to take the data dependency driven execution from the process local to the inter-
process level. The underlying halo exchange is implemented efficiently with GPI 2.0 (next generation
Global address space Programming Interface) which perfectly fits into this concept. GPI 2.0 is a
scalable communication API for one-sided, truly asynchronous communication primitives with remote
completion. It is the reference implementation of the GASPI specification and is released under
GPLv3 (www.gpi-site.com).

For the workshop, the strong scalability of a 15Hz single shot migration of the SEAM
dataset achieved by FRTM was analyzed. 
The SEAM dataset is a well established synthetic benchmark data set used
to evaluate the quality of seismic imaging algorithms. The computational cost for a single shot
migration is dominated by the finite difference modeling of the wave equation.

The underlying physical
model is given by a tilted transverse isotropic (TTI) medium. The finite difference approximation order
of the temporal discretization is second order, the spatial part is an 8-th order discretization.

Up to 512
nodes, the achieved scalability is almost perfect resulting in a parallel efficiency of 94\% at 512 nodes.
Beyond 512 nodes, one observes a drop in the parallel efficiency to 75\% at 1,024 nodes
and to 70\% at
1,536 nodes, respectively. This drop could be explained by the reduced inter-island network bandwidth
or by the transition from a 2D to a full 3D domain decomposition beyond 512 nodes.
The absolute FRTM run time for a single shot migration at 1,536 nodes is 54.5 seconds
(absolute single precision floating point performance of 210 TFlop/s), enabling interactive velocity
model building based on RTM as driving engine.

%\begin{figure}
%\centering\includegraphics[width=0.9\textwidth]{tti_speedup.pdf}
% \caption{
%Scaling of FRTM using GASPI on SuperMUC Phase 2.
% \label{fig:tti_speedup}
% }
%\end{figure}

\section{Gadget}

GADGET is a widely used, publicly available cosmological N-body / Smoothed
Particle Magnetohydrodynamics (TreePM-MHD-SPH) simulation
code. It uses an explicit
communication model implemented via MPI.

For gravity, GADGET uses a TreePM algorithm, based on the fully
MPI-parallelized FFTW library to perform the PM part of the gravity
solver. Thereby the long range gravity part is computed by sorting
the particles onto a mesh and solving Poisson’s equation via FFT; the
short range part is determined from a Tree-walk algorithm, computing a
direct sum of the forces between the particles and tree
nodes (TreePM). Hydrodynamics is solved via the Smoothed Particle
Hydrodynamics (SPH) method, which also make use of the Tree-walk
algorithm to perform the neighbor search needed. Additional processes
rely on various different numerical methods, for example transport
processes are solved by a conjugated gradient method. These different
physical modules are already optimized for mixed shared/distributed
memory architectures, making use of OpenMP. In addition most of the
physics modules -- for example star formation, thermal conduction,
black hole treatment and on-the-fly post-processing, which are
essential for modern cosmological applications -- have been prepared
for the latest architectures,
resulting in a 30\% performance improvement and efficient scalability
up to 131,072 cores on SuperMUC.

During the workshop, GADGET was used to perform one of
the largest cosmological hydrodynamical simulations (in terms of
resolution elements and covered cosmological volume) to date. The simulation followed
$2\times4,526^3$ particles and contained a detailed description of various,
complex, non-gravitational, physical processes
which determine the evolution of the cosmic baryons and impact their
observational properties (for details, see www.magneticum.org). 
Amongst them are the star formation and related
feedback; chemical pollution by SN Ia (Supernova Type Ia), SN II (Supernova
Type II) and asymptotic giant branch (AGB) winds; transport processes
like the thermal conduction as well as the evolution of black holes and
their related active galactic nucleus (AGN) feedback. All these processes
are self-consistently coupled with the underlying hydrodynamics.
GADGET also uses a new optimization of the tree walk (GreenTree)
as well as OpenMP lock improvements.

Including hyper-threading, GADGET
was running with 172,032 threads in an OpenMP/MPI hybrid configuration.
GADGET showed an extremely good I/O performance and reached 130Gbyte/s
writing and 150Gbyte/sec reading for the regular check pointing every 2h (each
having a total size of 66 TB). Overall, more than 4 Pbyte of data were produced.

\section{BQCD}

BQCD (Berlin Quantum Chromodynamics) has implemented various communication methods \cite{BQCD}:
MPI, OpenMP, MPI+OpenMP, as well as SHMEM (single sided communication).
The pronounced kernel is an iterative solver of a large system of linear equations (conjugate gradient).
Depending on the physical parameters, more than 95\% of the execution time is spent in the solver.
The dominant operation in the solver is a matrix-vector multiplication of a large and sparse hopping matrix.  The entries in a row are the eight nearest neighbours of one side of the four-dimensional lattice.
The conjugate gradient solver is communication intensive and represents the overall performance of the program in practical simulations. It serves as a good test to check the latency and resulting communication overhead.

\begin{figure}%
\includegraphics[width=0.8\textwidth]{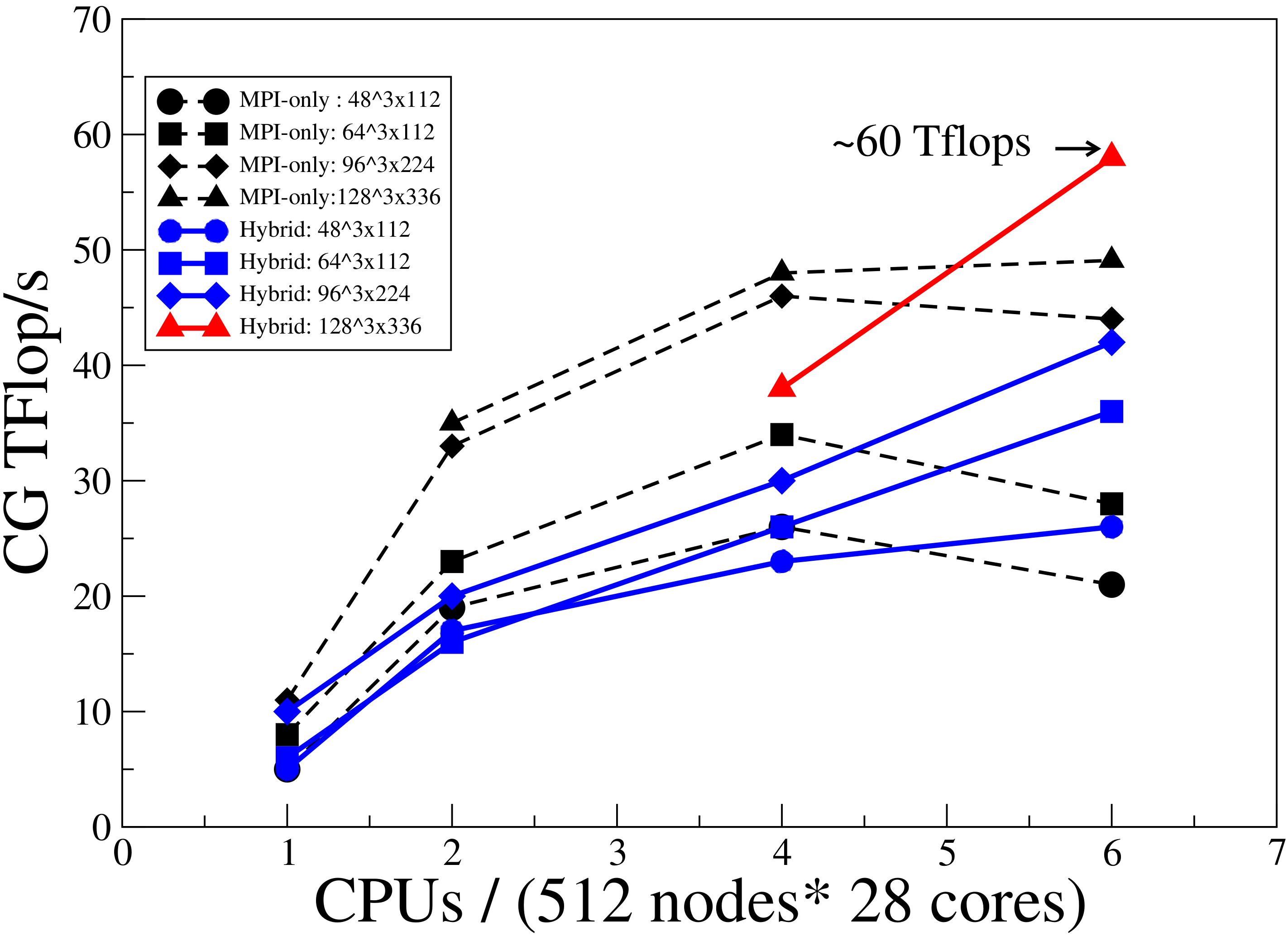}%
\caption{Strong scaling results of conjugate gradient solver for BQCD using MPI only and combination of MPI + 7 OpenMP threads per task, for different lattice sizes on SuperMUC Phase 2 system.
}%
\label{fig:BQCD}%
\end{figure}

Several strong and weak scaling runs were performed. One difficulty is to find the right lattice size to fit in to the data cache, another is to describe the full system with a local lattice which fits the Infiniband network.
Figure \ref{fig:BQCD} shows the performance results of the conjugate gradient solver for the pure MPI
versus the hybrid version, using lattice sizes of $48^3\times112$, $64^3\times112$, $96^3\times224$,
and $128^3\times336$. Up to 4 islands (57,344 cores), the best performance is achieved using MPI only.
On the full system, the hybrid version of the code delivers the best performance for all the lattice sizes using 7 OpenMP threads per MPI task.

%\begin{figure}
%\centering\includegraphics[width=0.6\textwidth]{bqcd_fig.png}
% \caption{
%Strong scaling results of conjugate gradient solver for BQCD using MPI only and combination of MPI + 7 OpenMP threads per task, for different lattice sizes on SuperMUC Phase 2 system.
% \label{fig:bqcd_fig}
% }
%\end{figure}

\section{CIAO}

CIAO is an in-house code developed in collaboration between the Institute for Combustion Technology at RWTH Aachen University and Sogang University. Depending on the physical nature of the problem, the Navier-Stokes equations are solved for either fully compressible flows or the low Mach limit. Various models for complex multi-physics flow configurations are available. For local mesh refinement or very stiff problems, a compressible multi-domain solver has been developed. CIAO is a structured, arbitrary order, finite difference code, which is parallelized with MPI. Use of spatial and temporal staggering of flow variables increases the accuracy of stencils with given size. The low Mach solver uses Crank-Nicolson time advancement and an iterative predictor corrector scheme. The compressible solver uses a low-storage five- stage explicit Runge-Kutta time advancement scheme. Scalar equations are discretized with a higher order WENO scheme, while momentum equations are spatially discretized with an arbitrary order central scheme. For Large-Eddy simulations, all of the subfilter stresses and scalar fluxes are modeled with dynamic Smagorinsky-type models using Lagrangian averaging along fluid particle trajectories. 
For the scale-out tests, a large-eddy simulation of a periodic channel with 20 additional scalars was performed using the compressible Navier-Stokes solver of CIAO which does not use any third-party libraries. 
The code shows good strong scaling behaviour from 7,168 up to 86,0168 cores.

%\begin{figure}
%\centering\includegraphics[width=0.6\textwidth]{ciao_fig1.jpg}
% \caption{
% Scaling of the CIAO code.
% \label{fig:ciao_fig1}
% }
%\end{figure}

\section{Results and Conclusions }

The workshop at LRZ showed that preparation of a simulation campaign is crucial for the success of the project. All aspects like scaling tests, choice of OpenMP/MPI balance, interval for checkpoint and restart files, good preparation of input files, I/O strategy, and risk management have to be addressed.
Under these conditions it was possible to use a brand new system like SuperMUC Phase 2 directly after installation and obtain scientific results right from the start.

One big advantage of the extreme scale-out workshop was that only one code was running at a time and this code was filling up the whole system. Thus, hardware bugs were much easier to detect and resolve. One especially hard to find bug was a combination of two timeouts and a hardware problem. During normal user operation this error would have been close to impossible to detect because of the low probability of three errors occurring simultaneously for smaller jobs.

It also became obvious that MPI is at its limits. The size of the MPI stack is growing on each node and for a system of almost 100,000 cores it occupies a significant amount of  memory. The startup time can exceed the range of minutes and become a significant part of the overall run time. One way to overcome this bottleneck is the use of hybrid OpenMP/MPI programming models. However, this implies very deep system knowledge on the user side, since process pinning and the choice of the OpenMP/MPI balance has to be evaluated and decided by the user. Furthermore, I/O strategies have to be developed and tested before the complete system can be used.
In the future, I/O libraries which can mediate this task become more and more important.
Even for hybrid openMP/MPI set-ups with a single MPI-task per node, problems arise due to internal limit of the MPI send/receive buffer. This limit is caused by the Integer*4 Byte implementation of the MPI index values. Such problems can be overcome by using application internal buffering.

\bibliography{parco_2015_extreme_scale_shorter}{}
\bibliographystyle{plain}

\end{document}